# Surface plasmon polariton assisted optical switching in noble bimetallic nanoparticle system

Sandip Dhara,[1,2,*] C.-Y. Lu,[2] P. Magudapathy,[3] Y.-F. Huang,[2,4]

W.-S. Tu,[2] K.-H. Chen[2,4,*]

[1] Surface and Nanoscience Division, Indira Gandhi Centre for Atomic Research, Kalpakkam-603102, India.

[2] Institute of Atomic and Molecular Sciences, Academia Sinica, Taipei 106, Taiwan

[3] Materials Physics Division, Indira Gandhi Centre for Atomic Research, Kalpakkam-603102, India.

[4] Center for Condensed Matter Science, National Taiwan University, Taipei 106, Taiwan

*Abstract*

Photoresponse of bimetallic Au-Ag nanoparticle embedded soda glass (Au-Ag@SG) substrate is reported for surface plasmon assisted optical switching using 808 nm excitation. Au-Ag@SG system is made by an ion beam technique where $Ag^+$ is introduced first in the soda glass matrix by ion exchange technique. Subsequently 400 keV $Au^+$ is implanted in the sample for different fluences which is followed by an ion beam annealing process using 1 MeV $Si^+$ at a fixed fluence of 2E16 ions.cm$^{-2}$. Characteristic surface plasmon resonance (SPR) peaks around 400 and 550 nm provided evidence for the presence of Au and Ag nanoparticles. An optical switching in the Au-Ag@SG system with 808 nm, which is away from the characteristic SPR peaks of Ag and Au nanoparticles, suggests the possible role of TPA owing to the presence of interacting electric dipole in these systems. The role of surface plasmon polariton is emphasized for the propagation of electronic carrier belonging to the conduction electron of Au-Ag system in understanding the observed photoresponse. Unique excitation dependent photoresponse measurements confirm the possible role of TPA process. A competitive interband and intraband transitions in the bimetallic system of Au and Ag with may be primarily responsible for the observation, which are validated qualitatively using finite difference time domain calculations where inter-particle separation of Au and Ag play an important role. Thus, a smart way of optical switching can be envisaged in noble bimetallic nanocluster system where long wavelength with higher skin depth can be used for communication purpose.

\* To whom corresponding should be addressed Email: dhara@igcar.gov.in; chenkh@pub.iams.sinica.edu.tw

The explosive progress in nano-optics is based on the nanoscale local field, which is greatly enhanced due to the resonant properties of metal nanoparticles.[1-4] The strong nanolocalized optical field induces enhanced nonlinear-optical phenomena and exhibits various prospective applications. Collective oscillation of conduction electrons, particularly in noble metal nanoclusters, with the excitation of visible light gives rise to surface plasmon polaritons (SPP) which propagate near the metal-dielectric interface.[5-7] It allows the observation of interference,[8,9] and imaging at the nanoscale.[10-12] The plasmon coupling within arrays of metal nanoparticles can lead to the formation of nanoscale hot spots in which the intensity of light from an incident beam can be concentrated by more than four orders of magnitude. The effect of light concentration by means of plasmon is most apparent in phenomena those are nonlinear in light intensity, as demonstrated recently by the on-chip generation of extreme-ultraviolet light by pulsed laser high harmonic generation.[13] This opens a wealth of prospects in lithography or imaging at the nanoscale through the use of soft x-rays. In fact, surface plasmon resonance (SPR) induced localized field generated second harmonic generation (SHG) using Au nanoparticles are well reported.[14-16] Among many such dramatic phenomena and applications, the most well known is the giant surface-enhanced or tip (metal coated) enhanced Raman scattering (SERS or TERS) that allows observation of single molecules.[17]

Understanding the interaction of light with matter at nanometre scale is a fundamental issue in optoelectronics and nanophotonics which are prerequisites for advanced sensor applications, namely, SPR assisted photoresponse manifested in the form of optical switching devises using all-optical signal processing,[2] bio-organic sensing beyond the sub-wavelength diffraction limit.[18] two photon absorption (TPA) enhanced SHG,[19,20] renewable energy resourcing by guiding and localizing light and considerable reduction in absorption layer thickness,[4] and medical therapy encompassing early detection of beginning of life (standard pregnancy test based on shift in plasmonic resonance frequency in the presence motion of molecular motors).[18] It is also used in controlling cancer cell by generating heat using the adiabatic heating of plasmonic vibration.[21] SPR in bimetallic nanocluster of noble metals of Au and Ag has sufficient interest for tuning the SPR peak position useful for various applications.[22-24] Among many direct applications of these bimetallic systems, SHG drew lot of attention in the Pt/Cu and Ag/Cu systems.[25,26]

We report here optical switching of a noble bimetallic system of Au and Ag nanoclusters embedded in glass matrix with excitation far beyond the plasmon resonance wavelength, confirming role of possible TPA related phenomenon in these photoresponse process. A bimetallic system of Au and Ag with competitive interband and intraband transitions may be primarily responsible for the observation. The observed phenomenon agrees very well qualitatively with the finite difference time domain (FDTD) calculations.

Bimetallic system of Au and Ag nanoparticles embedded in soda glass (Au-Ag@SG) were prepared in an ion beam technique where Ag is introduced in the ion exchange process, followed by implantation of atomic Au in the same matrix. At first, $Ag^+$ was introduced into soda glass (Si 21.49%, Na 7.1%, Ca 5.78%, Mg 0.34% and Al 0.15% by weight) by ion exchange technique. $Ag^+$-$Na^+$ ion exchange was performed by dipping soda lime glasses in a molten salt bath of $AgNO_3$ (0.2 N) in $NaNO_3$ in the temperature range of 320-350 $^o$C for few minutes with constant stirring in order to assure uniform nucleation of Ag clusters.[27] These samples were irradiated with 400 keV $Au^+$ for various fluences in the range of 2E16-5E16 ions.cm$^{-2}$ at room temperature at a pressure of 5E-7 mbar with current density around 1μA.cm$^{-2}$ to avoid heating for the nucleation of Au clusters. At this stage few atomic clusters of Au and Ag are formed (Fig. 1(a)).[28] These samples were finally ion beam annealed using 1 MeV $Si^+$ with high electronic energy loss at a fluence of 2E16 ions.cm$^{-2}$ to grow the Au and Ag nanoparticles. High energy ion beam with high electronic energy loss or electron beam annealing is a well established technique for both cluster growth.[28,29] The maximum range of Au and Si ions is calculated using SRIM and found to be 150 and 1600 nm,[30] respectively from the surface glass substrate as shown schematically in Fig. 1(a). We may mention here that Si will be in the form of nanoclusters constituting of few atoms having amorphous in nature and is less likely to contribute in the electrical conduction process. We may like to state here that electrical conduction in the metal ion implanted glass is highly unlikely with less than ppb level inclusion of Au and Si of ~1E16 ions.cm$^{-2}$ ranging over 1600 nm. Transmission electron microscopic (TEM) image shows (Fig. 1(b)) formation of nanoparticles with diameter around 5 nm. Selected area electron diffraction studies (outset Fig. 1(b)) show ring pattern showing all possible orientation in these nanoparticles. The rings correspond to (220), (111), (200) and (311) planes of Au (JCPDS #04-0784) and Ag (JCPDS #04-0783) phases and they are indistinguishable from each other as the *d*-spacing for both the phases are very close. For the confirmation of the presence of both Au and Ag, UV-visible absorption studies were carried out to show (Fig. 1(c)) peaks around 400 nm and 550 nm corresponding to Ag and Au nanoparticles, respectively. SPR absorption peaks corresponding to Au and Ag nanoparticles are reported to be present ~550 nm (2.25 eV),[31] and ~ 400 nm (3.1 eV),[27] respectively for the samples implanted with Au$^+$ at different fluences. Peaks corresponding to both Au and Ag nanoparticles are most prominent in the sample irradiated 5E16 ions.cm$^{-2}$ and we have used the sample for most of our present studies.

Au-Ag@SG sample was connected with sputter coated Au contact pad at two opposite edges of a maximum 10(length)x5(width) mm sample for photoconductivity studies, as shown in the schematic (supplementary information Fig. S1).[32] Photoconductivity studies were carried out using continuous wave (CW) diode lasers with emission length of 405 (80 mW), 532 (252 mW) and 808 nm (20 mW) using a probe station equipped with an optical microscope and two manipulators at room temperature. Laser spot was covering the area of both the contacts across the width. Under a magnified image in the microscope, each manipulator connected with a tungsten needle (tip size ≈2 μm) could be precisely controlled in their probing position on the micrometer-sized nickel electrodes of the device. A semiconductor characterization system (Keithley model 4200-SCS) was utilized to source the dc bias and to measure the current.

The FDTD simulations were performed using FDTD Solutions 7.0 from Lumerical Solutions. During the calculations, electromagnetic pulses in the wavelength of 405, 532 and 808 nm was launched into a box containing Au and Ag nanostructures for different inter-particle separation (2-10 nm with a step size of 2) to simulate a propagating plane wave interacting with the nanostructure. The metal nanostructure and its surrounding space were divided into 0.5 nm meshes. The refractive index (*n*) of the glass medium was taken to be 1.52.

Photoresponse measurement in bimetallic Au and Ag nanoparticles embedded in glass matrix was conducted alternatively under light illumination of different wavelengths and dark condition. Excitations with 405, 532 and 808 nm CW lasers were studied to show changes in the current upon the exposure of light (Fig. 2). Surprisingly, no photoresponse was observed for 405 nm excitation, which corresponds closely to the SPR peak position for Ag ~ 400 nm (Fig. 1(c)). So it was expected a strong SPR coupling with the incident light of 405 nm will influence the photoresponse characteristics. On the other hand, a change in measured current of 42% and 82% was observed for 532 and 808 nm light exposure (Fig. 2). Similar observation was also made for samples grown with $Au^+$ implantation of 2E16 and 3E16 ions.cm$^{-2}$ (inset in Fig. 2). The observed photocurrent may be understood in terms of SPP playing a role in the transport of conduction electrons belonging to Au-Ag system as carrier for the electrical conduction process. Again photoresponse of Au-Ag@SG system is surprising for 808 nm exposure, as SPR peak belonging to either for Ag at ~400 nm and Au at ~550 nm (Fig. 1(c)) fails to match with the excitation wavelength. We can also observe that the change in current is almost twice for 808 nm excitation than that observed for the exposure with 532 nm (Fig. 2) indicating a possible TPA behaviour for the former excitation. Unlike the conventional wisdom, the effect of TPA is realized in its manifestation of doubling the photocurrent with the excitation wavelength of 808 nm as compared to 532 nm excitation where a single photon is involved for the SPR assisted photoresponse. It may be relevant to mention that the SPP assisted generation of photocurrent does not occur across the band so a quadratic increase in the current cannot be expected. Instead the photocurrent is due to the transport of conduction electrons belonging to Au-Ag system as carrier for the electrical conduction process. So the role of TPA is limited to the coupling the conduction electrons of Au-Ag system. A competitive inter- and intraband transition in Au and Ag may be responsible for the observation. In order to understand it, FDTD calculations are analyzed showing no electromagnetic coupling for Au-Ag@SG system for 405 nm excitation even for an inter-particle separation between Au and Ag nanoparticles of 2 nm (Fig. 3), where $k$ is the electro-magnetic wave propagation vector, $E$ is the electrical field vector and intensity bars indicate $|E|^2$. A very strong dipole coupling, however is observed for both 532 nm and 808 nm excitations for the same system. It is well known that Au have a non-zero interband contribution and intraband transition alone contributes to the electro-magnetic light absorption process in Ag.[33-35] A bimetallic system of Au and Ag with competitive interband and intraband transitions, thus, may be primarily responsible for the observation. We may like to state here that, in the absence of proper diagnostic tool to identify Au and Ag separately we can safely assume that we have similar distribution for Au and Ag as shown in Fig. 1(b). Calculated dipole interaction of Au and Ag nanoparticles with different inter-particle separation is shown in the supplementary information (Figs. S2-S5).[32] Dipole interaction between Au and Ag nanoparticles was still active up to a gap of 4 nm for both 532 nm and 808 nm excitations (Fig. S2). The inter-particle separation of the nanoparticles, however are found to be lower than 4 nm in our study (Fig. 1(b)) to enunciate the SPP assisted transport of conduction electrons belonging to Au-Ag system as carrier for the electrical conduction process. The interactions die out with gap above 4 nm (Figs. S2-S5). Thus, it can be stated that the optical switching in the present system is essentially due to plasmon coupled TPA assisted SHG aiding in the electrical conduction process. With low absorption of 808 nm in the bimetallic system, the role of larger layer being involved in the electrical conduction process and increasing the photocurrent to almost twice the value of that observed in for 532 nm exposure is ruled out in the presence of strong coupling of electromagnetic wave of both 532 and 808 nm in the Au-Ag system (Fig. 3 and Fig. S2). The observation suggests possibility of optical switching in noble bimetallic nanocluster system where long wavelength with higher skin depth can be used for communication purpose.

In conclusion optical switching in the Au-Ag system using 808 nm excitation, which is away from surface plasmon resonance peaks of Au and Ag nanoparticles, suggests the presence of two photon absorption assisted processes owing to the presence of electric dipole interaction in these systems. Considering the photoresponse as an indirect manifestation of two photon absorption (TPA) process, doubling of excitation dependent photocurrent when excited with 808 nm as compared to that for 532 nm supports essentially the TPA model. A competitive interband and intraband transitions process in the bimetallic system of Au and Ag may be primarily responsible for the observation. Finite difference time domain calculations also support qualitatively the formation of interacting dipole for 532 and 808 nm excitations in these systems.

**Acknowledgements**

We are greatly in debt to L. C. Chen of CCMS, Taiwan for useful discussion.

**Supporting Information**

Photoresponse and FDTD studies for Au and Ag nanoparticles with different inter-particle separation between them.

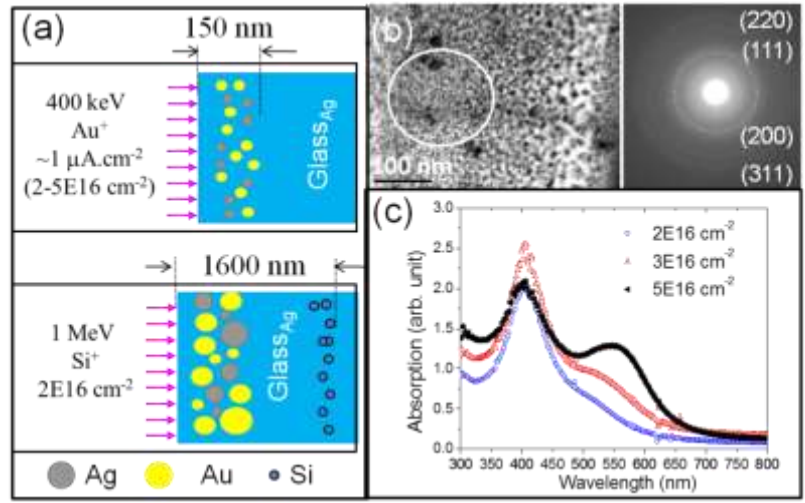

**FIG. 1.** (a) Schematic of implantation process in preparing Au-Ag embedded in soda glass system, (b) transmission electron microscopic image of Au and Ag nanoparticles. Outset shows diffraction rings corresponding to Au and Ag phases in the selected area electron diffraction pattern corresponding to the encircled region of the micrograph. (c) UV-visible absorption spectra for the samples grown in the ion beam process for different Au$^+$ fluences showing surface plasmon resonance corresponding to Au and Ag nanoparticles.

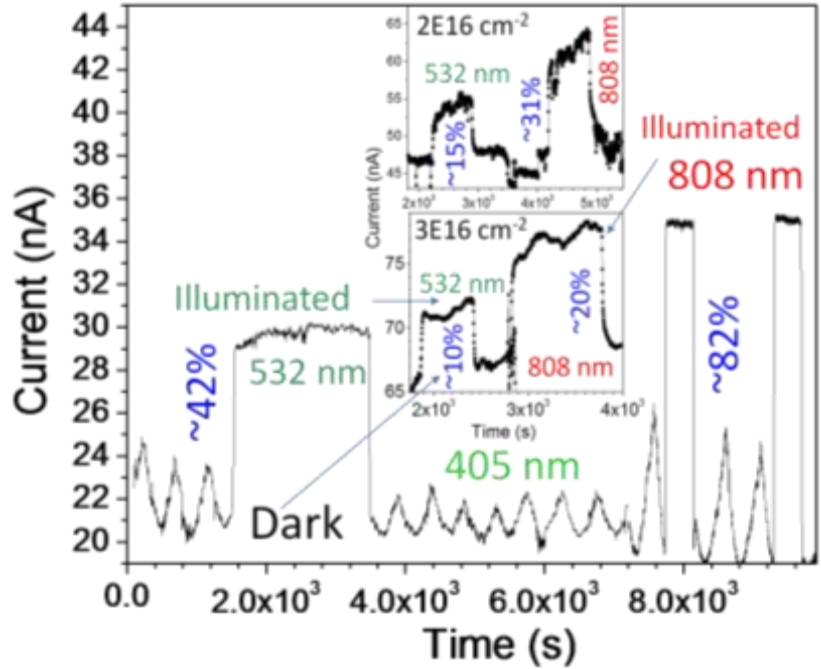

**FIG. 2.** Typical photoresponse of Au-Ag embedded in soda glass sample grown with $Au^+$ implantation at a fluence of 5E16 ions.cm$^{-2}$ with periodical dark and illumination to different laser wavelengths. Inset shows similar photoresponse of samples grown with $Au^+$ implantation at fluences of 2E16 and 3E16 ions.cm$^{-2}$. The studies show a double amount of change in current for 808 nm excitation than that observed for 532 nm exposure.

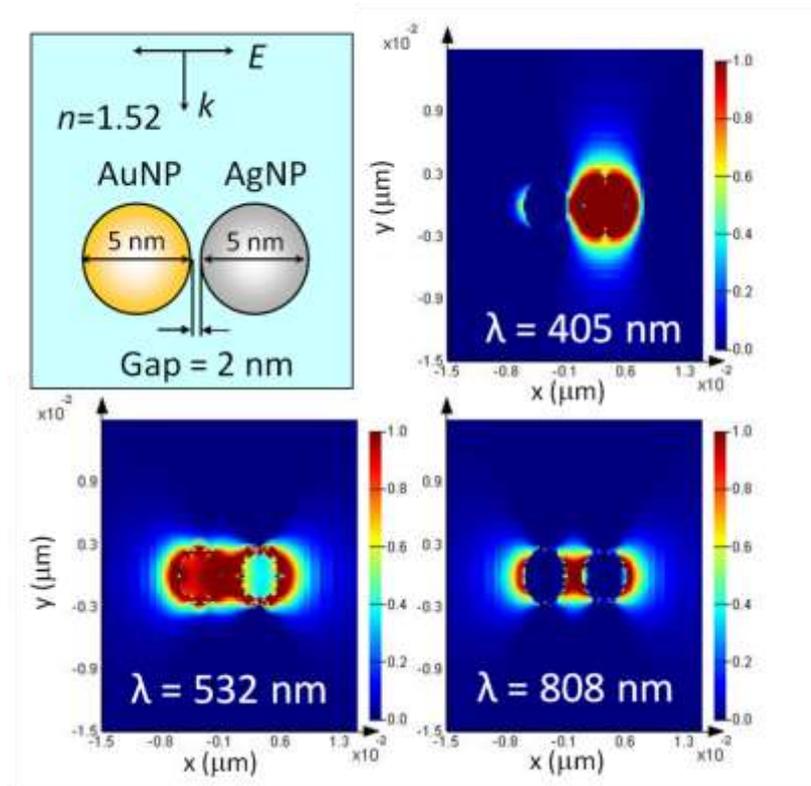

**FIG. 3.** Comparison of the electric field intensity enhancement contours for the interaction of electromagnetic radiation of different wavelengths with 5 nm Au and Ag nanoparticles separated by 2 nm in a medium with refractive index $n=1.52$. Here $k$ is the electro-magnetic wave propagation vector, $E$ is the electrical field vector and intensity bars indicate $|E|^2$.

**Supprting information :**

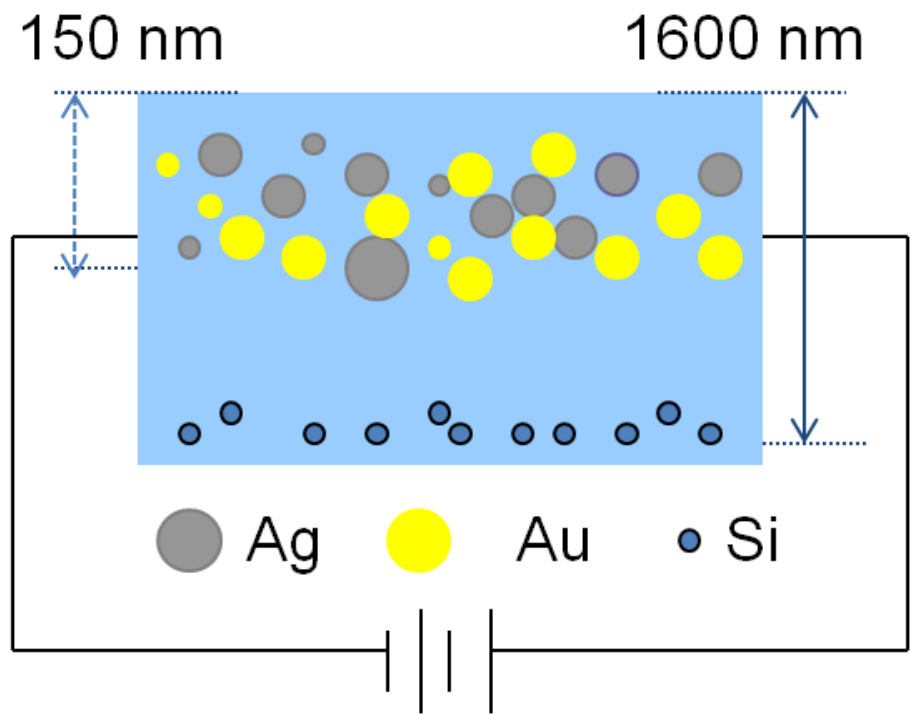

FIG. S1. Schematic of electrodes in the Au-Ag embedded in soda glass sample. As the resistances are very high ~ $10^6$ Ohm the two-point resistance will be independent of contact resistance and they are highly unlikely to change with exposure to light.

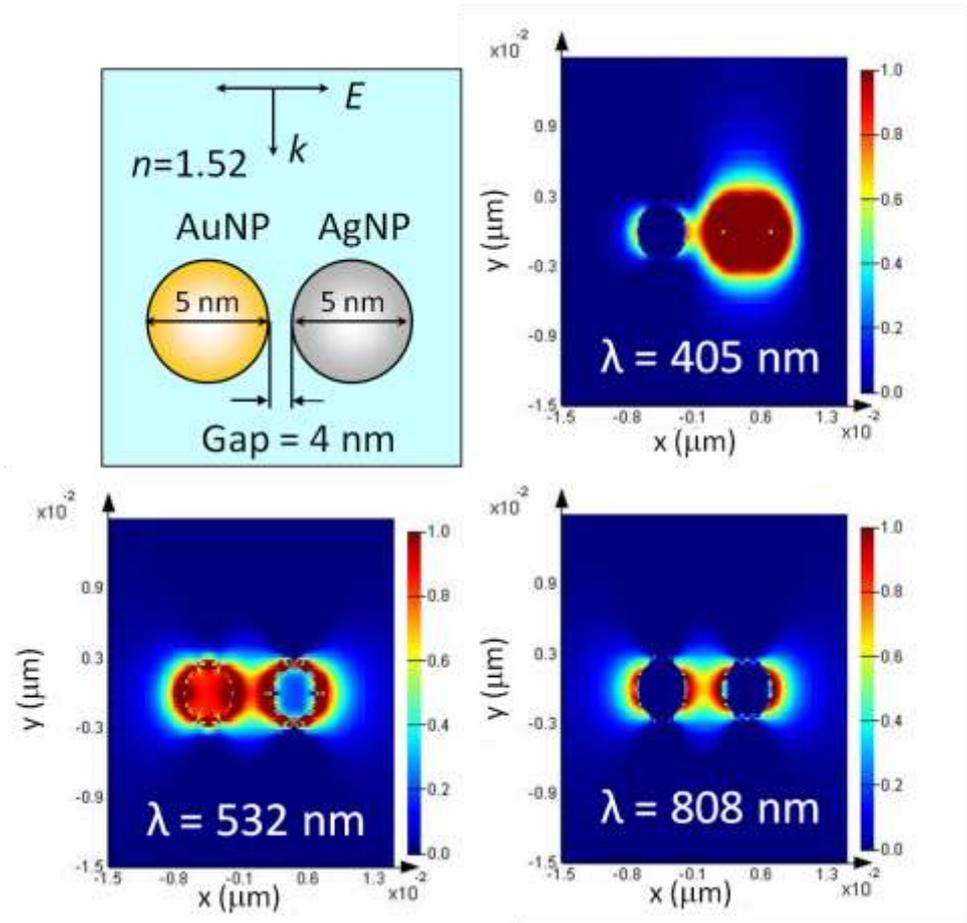

FIG. S2. Comparison of the electric field intensity enhancement contours for the interaction of electromagnetic radiation of different wavelengths with 5 nm Au and Ag nanoparticles separated by 4 nm in a medium with refractive index of $n=1.52$. Here $k$ is the electro-magnetic wave propagation vector, $E$ is the electrical field vector and intensity bars indicate $|E|^2$.

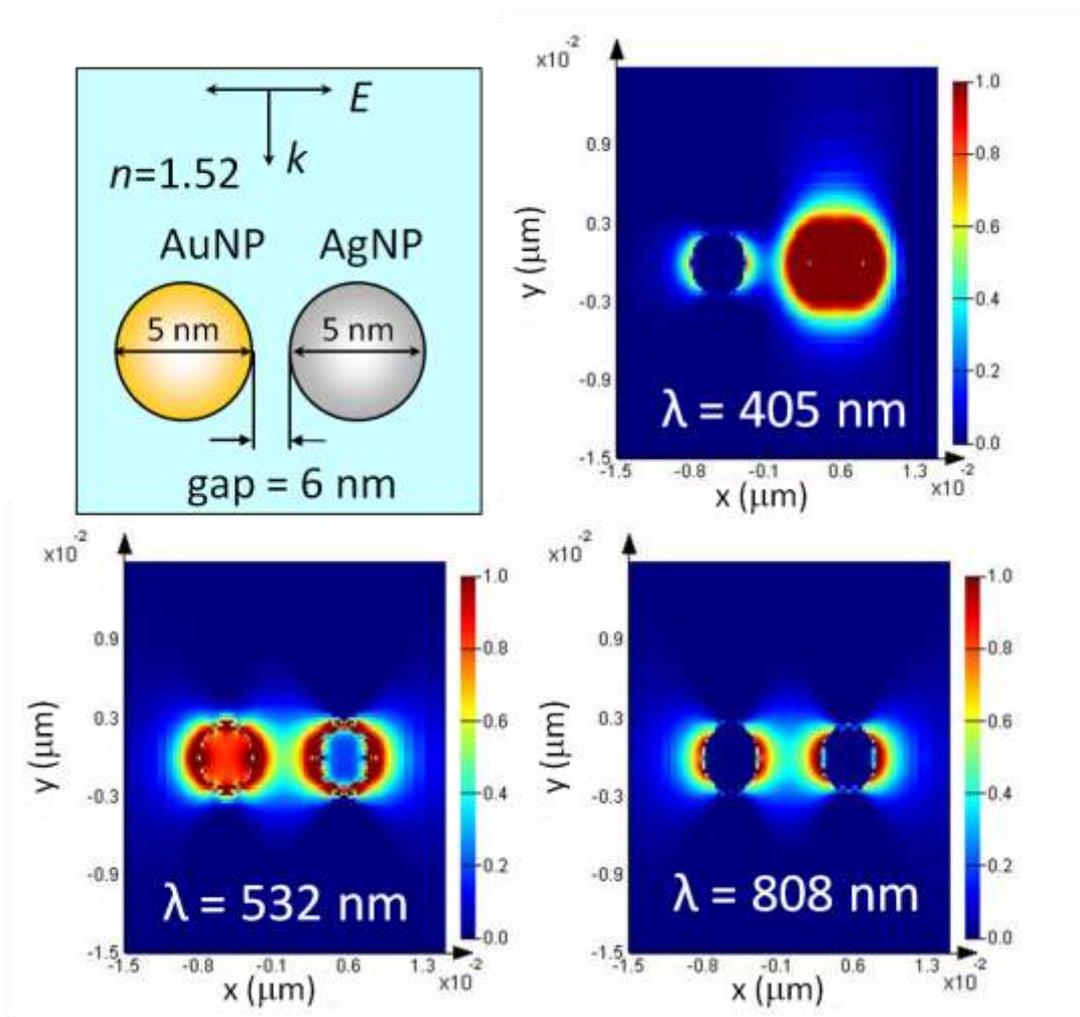

FIG. S3. Comparison of the electric field intensity enhancement contours for the interaction of electromagnetic radiation of different wavelengths with 5 nm Au and Ag nanoparticles separated by 6 nm in a medium with refractive index of $n=1.52$. Here $k$ is the electromagnetic wave propagation vector, $E$ is the electrical field vector and intensity bars indicate $|E|^2$.

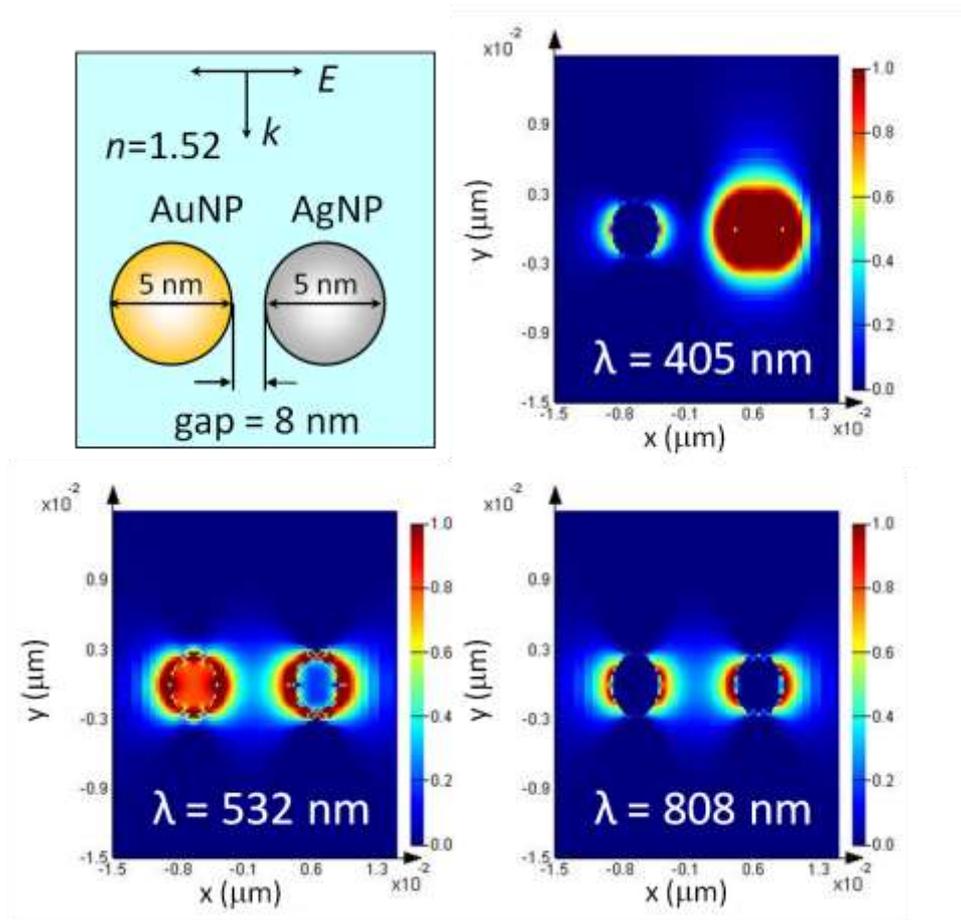

FIG. S4. Comparison of the electric field intensity enhancement contours for the interaction of electromagnetic radiation of different wavelengths with 5 nm Au and Ag nanoparticles separated by 8 nm in a medium with refractive index of $n=1.52$. Here $k$ is the electromagnetic wave propagation vector, $E$ is the electrical field vector and intensity bars indicate $|E|^2$.

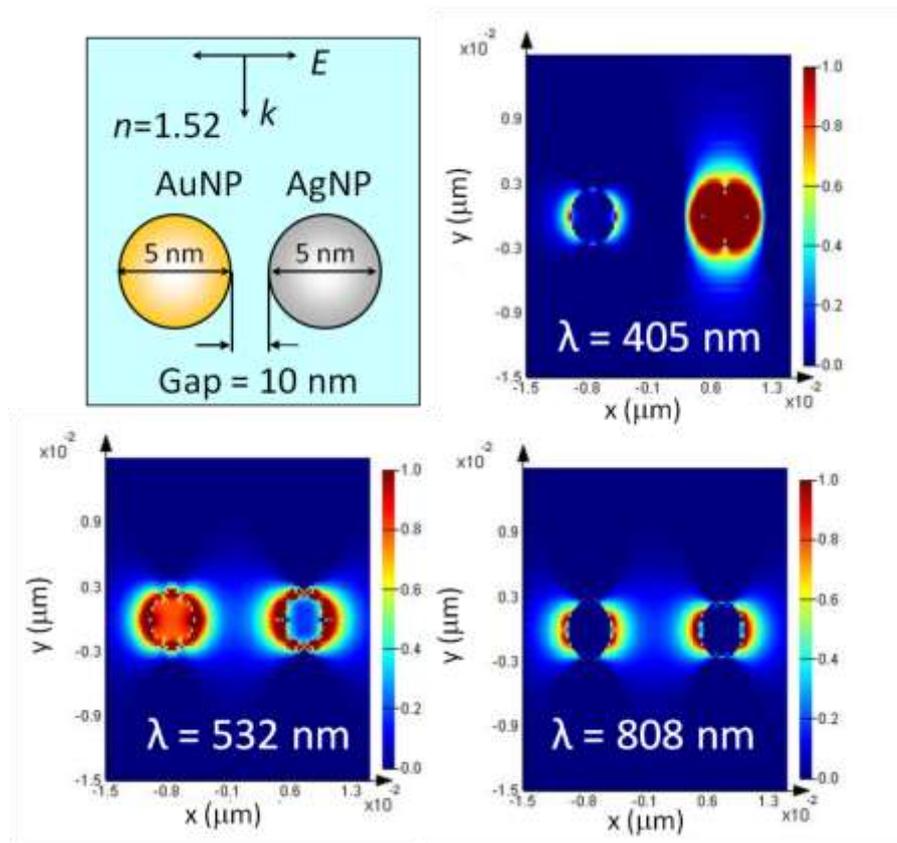

FIG. S5. Comparison of the electric field intensity enhancement contours for the interaction of electromagnetic radiation of different wavelengths with 5 nm Au and Ag nanoparticles separated by 10 nm in a medium with refractive index of $n=1.52$. Here $k$ is the electromagnetic wave propagation vector, $E$ is the electrical field vector and intensity bars indicate $|E|^2$.